\documentstyle[prl,aps,epsfig]{revtex}
\begin{document}

\title{London theory across superconducting phase transitions: application to UPt$_3$}
\author{D.F. Agterberg and Matthew J.W. Dodgson}
\address{Department of Physics, University of Wisconsin-Milwaukee, Milwaukee, WI 53201}
\address{Theory of Condensed Matter Group, Cavendish Laboratory, Cambridge, CB3 0HE, United Kingdom}

\maketitle
\begin{abstract}
For multi-component superconductors, it is known that the presence of symmetry breaking fields can lead to multiple superconducting phase
transitions.  This phenomenon is best illustrated in  UPt$_3$. Motivated by recent small angle neutron scattering experiments on the vortex state
of UPt$_3$, the London theory in the vicinity of such phase transitions is determined. It is found that the form of  this London theory is in
general quite different than that for conventional superconductors. This is due to the existence of a diverging correlation length associated with
these phase transitions. One striking consequence is that non-trivial vortex lattices exist arbitrarily close to H$_{c1}$. It is also found that
the penetration depth develops a novel temperature dependence and anisotropy. Results of this theory for UPt$_3$ are derived. Possible applications
to CeIn$_3$, U$_{1-x}$Th$_x$Be$_{13}$, electron doped cuprate superconductors, Sr$_2$RuO$_4$, and MgCNi$_3$ are also discussed.
\end{abstract}

\vglue 0.5 cm

One of the qualitatively new phenomena associated with unconventional superconductors is the existence of multiple degrees of freedom in the Cooper
pair wavefunction \cite{sig91,gor87}. This leads naturally to the possibility of phase transitions between superconducting states of different
symmetry. A well known example is UPt$_3$, which exhibits a phase diagram with three superconducting phases, each described by an order parameter
with different symmetry \cite{fis89,ade90,sau94}. Recently, using small angle neutron scattering (SANS),  Huxley {\it et al.} observed a
realignment of the flux-line lattice as the phase boundary between two different phases of UPt$_3$ is crossed \cite{hux00}. This measurement
provides the motivation to derive the London theory in the vicinity of such transitions (see Ref.~\cite{cha01} for a recent study of this problem).
Conceptually this raises the question: does the phase transition lead to non-trivial consequences for the low field vortex phases? Here we address
this question and find that the London theory indeed develops a novel form near the transition. The resulting theory profoundly changes the vortex
phase diagram from that expected from the usual London theory.  One result is the appearance of a rectangular flux lattice structure at the lower
critical field ($H_{c1}$). Such a lattice structure is not thought to be possible in the dilute vortex limit near $H_{c1}$. Physically, this arises
because the coherence length associated with the second transition diverges and thus becomes an important length scale near $H_{c1}$. The diverging
coherence length allows a non-trivial anisotropy to exist arbitrarily close to $H_{c1}$. This anisotropy dictates the appearance of the resulting
rectangular vortex phase. We show that such behavior should exist in UPt$_3$. Possible applications to U$_{1-x}$Th$_x$Be$_{13}$, CeIn$_3$, electron
doped cuprate superconductors, MgCNi$_3$, and Sr$_2$RuO$_4$ are also discussed.

\begin{figure}
\epsfxsize=2.0 in \center{\epsfbox{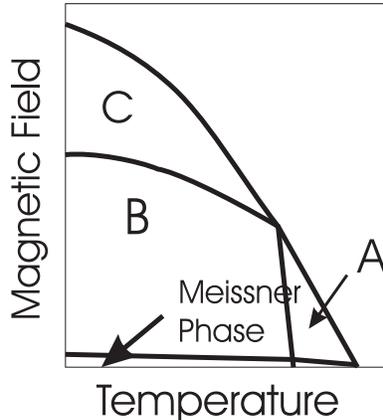}} \caption{Superconducting phase diagram of UPt$_3$. We are interested in the vortex phases near the
$A$ to $B$ phase transition for fields near $H_{c1}$.} \label{fig1}
\end{figure}

\noindent {\it Ginzburg Landau Free energy}

To demonstrate the physics described above the superconducting state of UPt$_3$ will be considered. The superconducting phase diagram of UPt$_3$
exhibits three distinct superconducting phases (see Figure 1). There have been many theoretical interpretations of this phase diagram (see
Ref.~\cite{sau94} for an overview). However, it appears that only those theories in which a doubly degenerate order parameter has its degeneracy
broken by a symmetry breaking field are consistent with the recent SANS measurements of Huxley {\it et al.}. In particular, it was observed that
the low-field flux-lattice in the $A$-phase is not aligned with any of the crystallographic axes  when the magnetic field is applied along the
six-fold symmetry axis. \cite{hux00}. This observation implies the existence of a two-component order parameter since it cannot occur for any of
the single-component representations of $D_{6H}$ \cite{cha01}.  The two most popular variants of this class of models are based on the $E_{2u}$ and
the $E_{1g}$ representations of the hexagonal point group \cite{sau94,par96}. The origin of the symmetry breaking field is taken to be
antiferromagnetic order \cite{sau94,par96}. However, the results are valid for any weak symmetry breaking field that reduces the $D_{6H}$ symmetry
to $D_{2H}$. We consider the $E_{2u}$ and the $E_{1g}$ theories with an applied magnetic field along the $c$-axis and ignore spatial variations
along this axis. The resulting free energy for the $E_{2u}$ or $E_{1g}$ representations of $D_{6H}$ is
\begin{eqnarray}
f=&-\alpha |\vec{\eta}|^2 +\gamma(\eta_+\eta_-^*+\eta_-\eta_+^*)+\frac{1}{2}\beta_1|\vec{\eta}|^4+
\beta_2|\eta_+|^2|\eta_-|^2 \nonumber\\
&\kappa|{\bf D}\eta_+|^2+|{\bf D}\eta_-|^2 \nonumber \\
& +\tilde{\kappa}[(D_-\eta_+)^*(D_+\eta_-)+c.c.) \label{gradeq1}
\end{eqnarray}
where $\alpha=\alpha_0(T_c-T)$, $D_j=\nabla_j-\frac{2ie}{\hbar c} A_j$, ${\bf h}=\nabla\times {\bf A}$, and ${\bf A}$ is the vector potential,  and
$D_{\pm}=D_x\pm i D_y$. This free energy neglects a term $h_z (|\eta_+|^2-|\eta_-|^2)$ which will not change any of the main physical results. This
term has been argued to be small \cite{par96}. Furthermore, there is a coupling of the symmetry breaking field to the gradient terms in the free
energy which has not been included \cite{sau94}. This coupling is required to give rise to the observed phase diagram at high fields \cite{sau94},
but does not result in any significant changes in the low-field regime discussed in this paper.  Experiments constrain the phenomenological
parameters for UPt$_3$. By fitting to the size of the specific heat jumps \cite{fis89} the value of $\beta_2/\beta_1=0.75$. From the splitting of
the two transition temperatures \cite{ade90} $\gamma/\alpha_0=17$ mK (this gives $T_{c,A}=500$ mK and $T_{c,B}=440$ mK). These parameters will be
used throughout this paper. The crucial parameter that drives the main results of this work is $\tilde{\kappa}$. The magnitude of
$\tilde{\kappa}/\kappa$ has been the subject of debate \cite{sau94,par96}. Sauls has argued that $\tilde{\kappa}/\kappa$ \cite{sau94} is small
while Joynt \cite{par96} has argued that $\tilde{\kappa}/\kappa$ may be as large as 1. In the following we will perform a perturbation in
$\tilde{\kappa}/\kappa$ and show that even a moderate $\tilde{\kappa}/\kappa$ has profound consequences on the vortex phase diagram for $T\approx
T_{c,B}$.

The goal of this work is to develop the London theory for the field along the six-fold symmetry axis as the $A$ to $B$ phase transition is crossed.
This problem has been addressed recently by Champel and Mineev in Ref.~\cite{cha01}. The result of this study is that the vortex lattice will be a
stretched hexagonal lattice near $H_{c1}$. This was found by determining the linear combination of $\eta_+$ and $\eta_-$ that stabilizes the
homogeneous ground state (this linear combination will be denoted below by $\psi_+$) and by examining the spatial variations of this order
parameter. Such an analysis will not result in the physics discussed in the introduction. This physics is found by including the order parameter
component orthogonal to that which minimizes the homogeneous minimizing free energy (denoted by $\psi_-$). The component $\psi_-$  is induced by
spatial variations of $\psi_+$ through the $\tilde{\kappa}$ term in the free energy. As will be shown below, this coupling qualitatively changes
physics of the London theory from that found by setting $\psi_-=0$.

The homogeneous free energy stabilizes $(\eta_+,\eta_-)=(1,1)$ near $T_{c+}$ (the $A$ phase) and at a lower temperature there is a second
transition to the $B$ phase, a state $(\eta_+,\eta_-)=|\eta|[\cos(\varphi/2),\sin(\varphi/2)]$ where $\varphi\ne \pi/2$ and varies with
temperature. This transition occurs when $\beta_2>0$. This free energy will be used to determine the London theory with the lowest order non-local
correction (similar to what was done for the superconducting state of  Sr$_2$RuO$_4$ \cite{hee99}). To implement this the $(\eta_+,\eta_-)$ is
rotated through an angle $\varphi/2$ to give a basis $(\psi_+,\psi_-)$. The angle $\varphi$ is chosen so that $(\psi_+,\psi_-)\propto(1,0)$ is the
stable solution for all temperatures. For $\gamma>0$, this requires that $\varphi=\pi/2$ in the $A$ phase and that $\sin \varphi =
\frac{2\gamma}{\beta_2 \alpha}$ and $|\psi_+|^2=\alpha$ in the $B$ phase. The rotated free energy is
\begin{eqnarray}
f=&-\alpha |\vec{\psi}|^2 +\gamma[\cos  \varphi (\psi_+\psi_-^*+\psi_-\psi_+^*)+\sin  \varphi(|\psi_+|^2-|\psi_-|^2)] +\frac{1}{2}\beta_1|\vec{\psi}|^4+ \nonumber \\
& \frac{1}{4}\beta_2[(|\eta_+|^2+|\eta_-|^2)^2 -[\cos  \varphi (\psi_+\psi_-^*+\psi_-\psi_+^*)+\sin  \varphi(|\psi_+|^2-|\psi_-|^2)]^2]\nonumber
\\& \kappa(|{\bf D}\psi_+|^2+|{\bf D}\psi_-|^2) \nonumber
\\ &
+\frac{1}{2}\tilde{\kappa}[[(D_-\psi_+)^*(D_+\psi_-)-(D_-\psi_-)^*(D_+\psi_+)+c.c] +\cos
\varphi[(D_-\psi_+)^*(D_+\psi_-)+(D_-\psi_-)^*(D_+\psi_+)+c.c] \nonumber
\\ &+\sin \varphi[(D_-\psi_-)^*(D_+\psi_-)-(D_-\psi_+)^*(D_+\psi_+)+c.c]] \label{eq2}
\end{eqnarray}
where $cc$ denotes complex conjugate. If $\tilde{\kappa}=0$ then the two components $\psi_+$ and $\psi_-$ are decoupled, implying that the solution
in the vortex phase near $H_{c1}$ is described by $\psi_-=0$. Once $\tilde{\kappa}$ becomes non-zero, this term induces a non-zero $\psi_-$ in the
vortex phase. We treat the term in $\tilde{\kappa}$ up to second order perturbatively. All terms in $O(\psi_-^3)$, $O(\tilde{\kappa}\psi_-^2)$ and
higher are ignored in the free energy which results in a quadratic free energy for $\psi_-$. The free energy is minimized for $\psi_-$ and the
solution is used to get an effective free energy for $\psi_+$. A subtle point is this derivation is the treatment of the relative phase between the
two components $\psi_+$ and $\psi_-$. Within the approximations made below it can be shown that this phase is spatially uniform. The effective free
energy for $\psi_+$ is then treated with standard methods to arrive at a London free energy.  In particular, writing
$\psi_+=|\psi_+|\exp^{i\theta}$, taking $|\psi_+|$ to be spatially uniform and equal to the homogeneous value, introducing ${\bf v}_S=\nabla
\theta-\frac{2 e}{\hbar c}{\bf A}$, keeping only terms of order $v_s^2$, minimizing this free energy with respect to ${\bf A}$ (to find ${\bf v}_S$
as a function of ${\bf b}=\nabla\times {\bf A}$) yields the following London energy for vortex lattice in the $A$ phase [the vortex lattice was
introduced as a gaussian source (with a cutoff given by the $A$ phase coherence length $\xi_{A,+}$) for the London equation, as was done in
Refs.~\cite{hee99,kog97}]
\begin{equation}
F=F_0+\frac{B_0^2}{8\pi}\sum_{{\bf q}={\bf G}} \frac{e^{-{\bf q}^2 \xi_{A,+}^2}}{\Big[
1+\lambda_A^2\big(\frac{\kappa}{\kappa-\tilde{\kappa}}q_x^2+\frac{\kappa}{\kappa+\tilde{\kappa}}q_y^2\big)+
\lambda_A^4\frac{\tilde{\kappa}^2}{\kappa^2}\frac{(q_x^2-q_y^2)^2}{\frac{\lambda_A^2}{\xi_{A,-}^2}+\lambda_A^2{\bf q}^2}\Big]} \label{eq3}
\end{equation}
where ${\bf q}$ is summed over all reciprocal lattice vectors of the vortex lattice, $B_0$ is the spatially averaged magnetic induction,
$\lambda_A=\frac{\beta_1+\beta_2/2}{\alpha+\gamma}\frac{\hbar^2 c^2}{32\pi e^2 \kappa}$ is the penetration depth in the $A$ phase when
$\tilde{\kappa}=0$, and $\xi_{A,-}=\frac{2\gamma\beta_1-\alpha\beta_2}{\beta_1+\beta_2/2}\frac{1}{\kappa}$ is the coherence length that describes
the transition to the $B$ phase from the $A$ phase. In the $B$ phase vortex lattice the London energy is
\begin{equation}
F=F_0+\frac{B_0^2}{8\pi}\sum_{{\bf q}={\bf G}}\frac{e^{-{\bf q}^2 \xi_{B,+}^2}}{ \Big[1+\lambda_B^2\big(\frac{\kappa}{\kappa-\tilde{\kappa}\sin
\varphi }q_x^2+\frac{\kappa}{\kappa+ \tilde{\kappa}\sin \varphi}q_y^2\big)+
\lambda_B^4\frac{\tilde{\kappa}^2}{\kappa^2}\frac{(q_x^2-q_y^2)^2+4\cos^2 \varphi q_x^2q_y^2}{\frac{\lambda_B^2}{\xi_{B,-}^2}+\lambda_B^2{\bf
q}^2}\Big]}
\end{equation}
where $\lambda_B=\frac{\beta_1}{\alpha}\frac{\hbar^2 c^2}{32\pi e^2 \kappa}$ is the penetration depth in the $B$ phase when $\tilde{\kappa}=0$,
$\sin \varphi=\frac{2\gamma\beta_1}{\alpha\beta_2}$ (note $\sin\varphi=1$ for $T=T_{c,B}$),
$\xi_{B,-}=\frac{\alpha\beta_2\cos^2\varphi}{\kappa\beta_1}$ is the coherence length that describes the transition to the $A$ phase from the $B$
phase, and $\xi_{B,+}$ is the coherence length associated with the $B$ phase $\Psi_+$ order parameter component. We take
$\kappa_{GL}=\lambda_B/\xi_{B,+}=60$ as is quoted by Huxley {\it et al.} \cite{hux00} and for simplicity we take
$\lambda_B/\xi_{B,+}=\lambda_A/\xi_{A,+}$ (which is correct to within 10\%).

The interesting physical behavior arises because $\frac{\lambda^2_{A(B)}}{\xi_{A (B),-}^2}=0$ at the $T_{c,B}$ so that the non-local term (the term
proportional to $[\tilde{\kappa}/\kappa]^2$) becomes of order $q^2$ and thus {\it as important as the local term}. This is due to the divergence of
the correlation length associated with the $A$ to $B$ phase transition.  Typically near $H_{c1}$ the only relevant length scale is the distance
between vortices and the usual analysis can be applied \cite{kog81}. However, near the $A$ to $B$ phase transition the diverging correlation length
appears as a second relevant length scale and, as will be seen below, plays an important role in the London theory.  Corrections beyond the mean
field model used here can be included in the London theory. This will change the temperature dependence of $1/\xi_{A (B),-}$, but will not change
the powers of $q$ that appear in the non-local term \cite{cha95}. It is the $q$ dependence of the London theory that is important in what follows.
The complete London theory will also contain additional non-local terms (like those that appear in the borocarbides \cite{kog97}), however these
corrections are small because they formally carry an additional factor $(\xi_{A(B),+}/\lambda_{A(B)})^2\approx 1/3600$ \cite{kog97,hux00}.
Consequently, near $H_{c1}$, the above London energy will describe UPt$_3$ very well in the vicinity of the $T_{c,B}$. Below we discuss some
observable consequences of this London free energy.
\\
\noindent{\it Penetration Depth}

The temperature dependence of the non-local term becomes of order $q^2$ at the $T_{c,B}$. This indicates that the temperature dependence of the
penetration depth will exhibit a non-trivial behavior. For the field along the $c$-axis and a surface with a normal along the crystallographic
${\bf a}$ axis, the penetration depth in the $B$ phase is given by
\begin{equation}
\lambda^2=2\lambda_B^2\frac{1+\frac{\tilde{\kappa}^2}{\kappa^2}}{1-\frac{\tilde{\kappa}}{\kappa}\sin\varphi+\frac{\lambda_B^2}{\xi_{B,-}^2}
+\sqrt{\big(1-\frac{\tilde{\kappa}}{\kappa}\sin\varphi-\frac{\lambda_B^2}{\xi_{B,-}^2}\big)^2-4\frac{\tilde{\kappa}^2}{\kappa^2}\frac{\lambda_B^2}{\xi_{B,-}^2}}}
\end{equation}
the result for the $A$ phase is given by changing the label $B$ to $A$ and setting $\sin \varphi=1$.  This behavior differs from the usual Ginzburg
Landau behavior. The penetration depth will also exhibit an in-plane anisotropy that is maximal at $T_{c,B}$. However, this anisotropy will be
obscured by the existence of multiple domains of a symmetry breaking field.
\\
\noindent{\it Vortex lattice structure}

\begin{figure}
\epsfxsize=1.5 in \center{\epsfbox{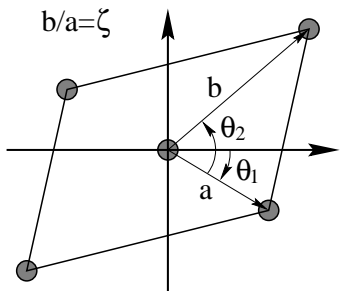}} \caption{Real space unit cell of the vortex lattice.} \label{fig2}
\end{figure}

The vortex lattice unit cell used here is defined in Fig.~\ref{fig2}.

\begin{figure}
\epsfxsize=3.5 in \center{\epsfbox{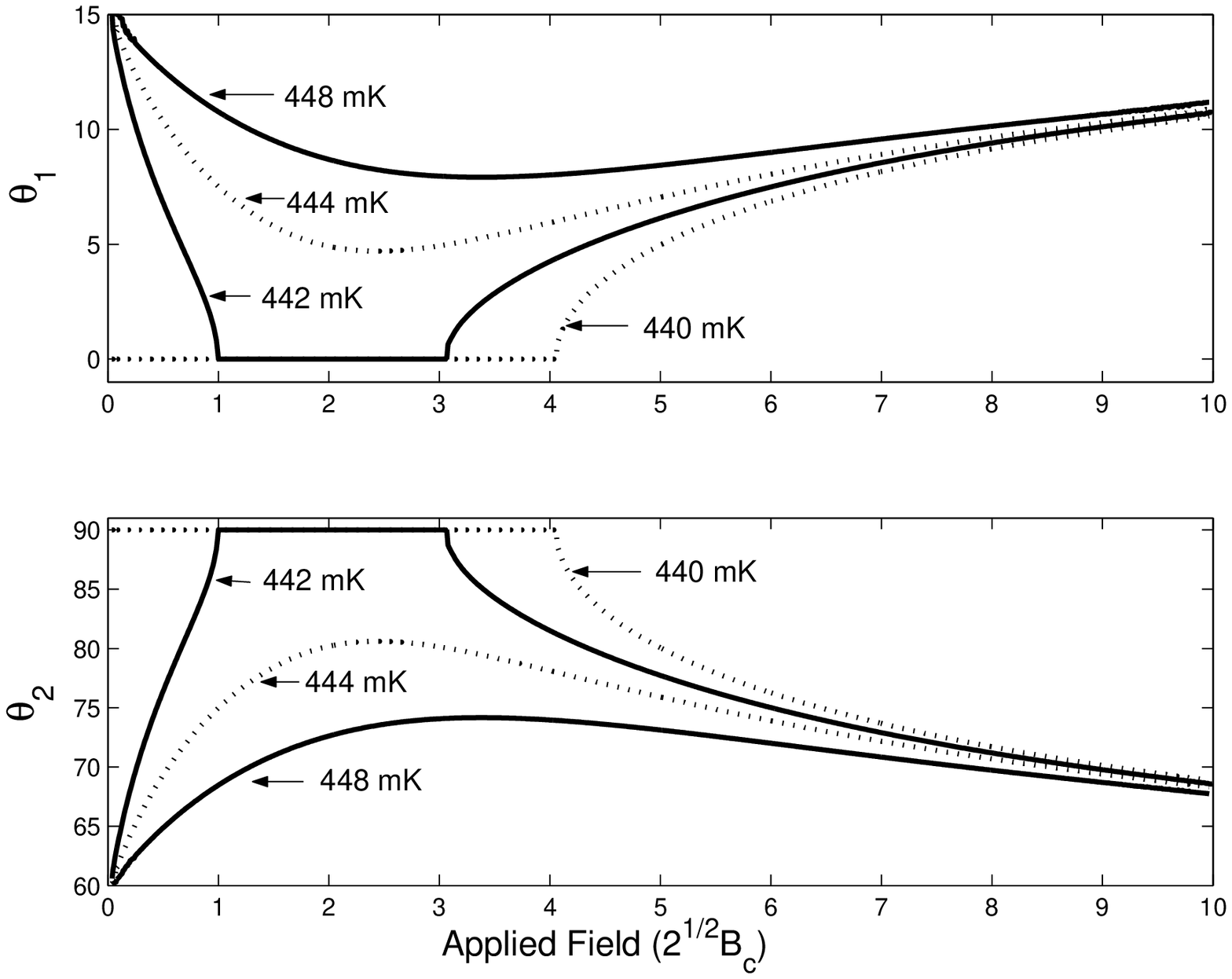}} \caption{Vortex lattice structure as a function of applied field and temperature in the $A$-phase
for temperatures near the $A$ to $B$ transition ($T_{c,B}=440$ mK). Results have been included for fields $H>0.1 \sqrt{2} B_c$. The parameter
$\zeta=1$ throughout the entire parameter range shown above.}\label{fig3}
\end{figure}

The London energy derived above has the most significant impact on the vortex lattice structures for temperatures near $T_{c,B}$. Earlier analysis
implied a stretched hexagonal lattice near $T_{c,B}$ \cite{cha01} which is not strongly field dependent (this was found for arbitrary
$\tilde{\kappa}/\kappa$). Such a lattice results from a re-scaling of the variables $q_x$ and $q_y$ to remove a $q_x^2-q_y^2$ orthorhombic symmetry
term in the London energy \cite{kog81}. The resulting London energy in the re-scaled variables is isotropic (which implies a hexagonal vortex
lattice in the re-scaled parameters).  This analysis neglects the divergence of $\xi_{A(B),-}$ as the temperature approaches $T_{c,B}$, which
changes the results significantly.   We performed a numerical analysis of the vortex lattice structure in the $A$ phase for
$\tilde{\kappa}/\kappa=0.36$, the results are shown in Fig.~\ref{fig3}. In this analysis, the $q_x^2-q_y^2$ anisotropy in the London energy is
removed by re-scaling the $q_x$ and $q_y$ \cite{kog81}. Since the re-scaled and original $q_x$ and $q_y$  differ by only a factor
$\tilde{\kappa}/\kappa$, this re-scaling is ignored in the non-local term (note including the re-scaling  in the non-local term is equivalent to
keeping factors $O(\tilde{\kappa}^3/\kappa^3)$ which have been previously thrown out).  Fig.~\ref{fig3} only show results for $H\ge0.1 \sqrt{2}B_c$
(numerically it was found that $H_{c1}=0.03 \sqrt{2}B_c$ note also $H_{c2}\approx 60 \sqrt{2}B_c$). As can be seen for $T=T_{c,B}=400$ mK, the
low-field vortex lattice phase is not hexagonal, but square for a substantial range of applied magnetic field. For $T\ne T_{c,B}$, the low-field
vortex lattice is hexagonal. This can be understood by noting that $\xi_{A,-}$ is finite for $T\ne T_{c,B}$ and consequently, only the usual London
theory applies at the lowest fields (since the vortices become infinitely separated as $H\rightarrow H_{c1}$). However for $T=T_{c,B}$, $\xi_{A,-}$
is infinite, so that the non-local terms in the London theory are important even as $H\rightarrow H_{c1}$. Furthermore, it was found that for the
lowest fields that were numerically accessible, the stable vortex phase for $T=T_{c,B}$ was rectangular (a square lattice stretched along one of
its sides). This lattice was found to be stable for $H_{c1}<H<\approx 1.003 H_{c1}$ (this field range is not shown in Fig.~\ref{fig3}). In
comparing the theory to existing experimental results, it is important to point out that the  high-field vortex phase agrees with that observed by
Huxley {\it et al.} \cite{hux00}. We found that the high-field (nearly) hexagonal lattice has one of the basis vectors rotated 15 degrees from the
crystallographic ${\bf a }$ axis which is in agreement Ref.~\cite{hux00}. This is a non-trivial result of the theory found here. In particular, the
sign of the non-local term derived in Eq.~\ref{eq3} is opposite that used by Huxley {\it et al.} to explain the observed orientation. For smaller
$\tilde{\kappa}/\kappa$, the vortex phase diagram is very similar to that found in Fig.~\ref{fig3} with one important difference: the region where
the square vortex lattice phase appears becomes suppressed to lower fields and to temperatures closer to $T_{c,B}$.

Given that the intriguing behavior found here was due the diverging correlation length at $T_{c,B}$, it is clear that this physics is not
restricted to UPt$_3$. It will arise in other multi-component superconductors that exhibit multiple phase transitions. Some additional  candidates
for this behavior are U$_{1-x}$Th$_x$Be$_{13}$ \cite{hef90}, Sr$_2$RuO$_4$ \cite{mae94,ric95,mae01}, CeIn$_3$ \cite{mat98}, electron doped cuprates
\cite{bis02}, and MgCNi$_3$\cite{he01,ros02}.  In U$_{1-x}$Th$_x$Be$_{13}$ multiple phase transitions have been observed \cite{hef90}. Cubic
CeIn$_3$ is likely to have a superconducting phase diagram similar to UPt$_3$ with a two-component $d$-wave order parameter \cite{agt02}.
Sr$_2$RuO$_4$ has been established as an unconventional superconductor with a two-component order parameter \cite{mae94,ric95,mae01}. The
application of uniaxial stress should give rise to multiple transitions in this case  \cite{sig87}. Electron doped cuprates have been argued to
exhibit a $d$ to $s$-wave pairing symmetry transition with doping \cite{bis02}. If this is indeed the case, then there should exist a particular
doping where the two symmetries co-exist. This would imply multiple phase transitions.  Finally, cubic MgCNi$_3$ has been argued to be a
spin-triplet superconductor \cite{ros02}, for which it is likely that a multi-component order parameter arises \cite{sig91}. In this case, the
application of uniaxial pressure will then be required to give rise to multiple superconducting phase transitions.

In conclusion, we have shown that for superconductors that exhibit multiple superconducting phases in zero applied magnetic field,  the London
theory in the vicinity of phase transitions takes on a very novel form. In particular, the diverging correlation length associated with the phase
transition allows non-trivial vortex lattice structures to exist arbitrarily close to $H_{c1}$.  It was also found that the penetration depth will
exhibit an unusual temperature dependence and anisotropy near such transitions. This theory was applied to UPt$_3$ and the possible application of
this theory to U$_{1-x}$Th$_x$Be$_{13}$, Sr$_2$RuO$_4$, CeIn$_3$, electron-doped cuprates, and MgCNi$_3$ was discussed.

We thank A. Huxley, R. Joynt, J. Moreno, P. Rodiere, and J. Sauls for useful discussions. DFA was supported by an award from Research Corporation
and by a Research Committee Award from the University of Wisconsin-Milwaukee. MJWD is supported by an EPSRC Advanced Fellowship AF/99/0725.

\end{document}